# Tenfold reduction of Brownian noise in optical interferometry

Garrett D. Cole[1,2,*], Wei Zhang[3,*], Michael J. Martin[3], Jun Ye[3], and Markus Aspelmeyer[1]

[1] *Vienna Center for Quantum Science and Technology (VCQ), Faculty of Physics, University of Vienna, A-1090 Vienna, Austria*

[2] *Crystalline Mirror Solutions GmbH, A-1090 Vienna, Austria*

[3] *JILA, National Institute of Standards and Technology and University of Colorado, Boulder, Colorado 80309-0440, USA*

**Thermally induced fluctuations impose a fundamental limit on precision measurement. Today's most advanced technologies for measuring time and space[1], prominently optical atomic clocks[2,3] and interferometric gravitational wave detectors[4], have now encountered this ultimate barrier. These systems rely on optical interferometry, where thermally-driven fluctuations result in modifications of the optical path length and hence in unavoidable "thermal noise"[5]. The need to minimize these effects has led to major advances in the design of optical cavities, which have continually redefined the state-of-the-art in optical precision sensing[6-11]. Still, the most significant impediment for achieving enhanced performance is the Brownian motion of the cavity's high-reflectivity multilayer coatings[10,12]. According to the fluctuation-dissipation theorem[13], this motion is directly linked to the mechanical damping in the constituent materials of the coating. Thus, the long-standing challenge has been to identify materials simultaneously capable of high reflectivity and low mechanical dissipation. Over the last decade, the mechanical damping**

---

[*] These authors contributed equally to this work.

**of high-reflectivity multilayers has at best been reduced by a factor of two[14]. Here we demonstrate a new paradigm in optical coating technology. Building upon advancements in semiconductor lasers[15], quantum optomechanics[16], and microfabrication[17], we have developed cavity mirrors comprising compound-semiconductor-based epitaxial multilayers, which exhibit both intrinsically low mechanical loss and high optical quality. Employing these "crystalline coatings" as end mirrors in a Fabry-Pérot cavity, we obtain a finesse of 150,000. More importantly, at room temperature, we observe a thermally-limited noise floor consistent with a tenfold reduction in mechanical dissipation when compared with the best dielectric multilayers. These results pave the way for the next generation of ultra-sensitive optical interferometers, as well as for new levels of laser stability.**

Optical precision measurement relies on the accurate sensing of changes in phase of a probe laser beam, which in turn requires ultra-stable optical interferometers. The best performance to date is achieved with mirrors implementing high-reflectivity multilayer coatings on transparent substrates, both of which exhibit sub-nanometer surface roughness. Dielectric $SiO_2/Ta_2O_5$ multilayers deposited by ion beam sputtering (IBS) have represented the state of the art in high-reflectivity optical coatings since the 1980s, with optical absorption at the parts-per-million (ppm) level[18]. This has led to enormous progress in a broad spectrum of applications including sub-attometer displacement sensitivities for gravitational wave observatories[4] and frequency stabilities at the $10^{-16}$ level for metrology applications[9-11].

In spite of their superior optical properties, the amorphous thin films at the heart of these coatings exhibit excess mechanical damping[19,20] driven by internal losses of the high-index tantala ($Ta_2O_5$) layers[21]. This results in significant displacement fluctuations of the mirror surface arising from thermally driven mechanical modes, typically referred to as "coating thermal noise"[22]. In general, the magnitude of such Brownian motion can be quantified via the frequency-dependent noise power spectral density (NPSD), $S_x$, and depends strongly on the mechanical damping in the material system. Specifically, $S_x \propto k_B T \cdot \phi \cdot 1/f$ ($k_B$: Boltzmann constant, $T$: temperature, $f$: Fourier frequency), where the mechanical loss angle $\phi$, given by the imaginary component of a complex Young's modulus $E(\omega) = E_0[1 + i\phi(\omega)]$, characterizes the mechanical energy dissipation rate. The overall Brownian noise floor is determined by the loss angle contributions of all cavity components, namely the substrate $\phi_{sub}$, spacer $\phi_{spacer}$, and coating $\phi_c$ [12]. This is the essence of the fluctuation-dissipation theorem in its application to optical interferometry. An independent source of significant coating-related noise arises from thermal fluctuations in the multilayer and substrate driven by the finite thermal expansion coefficient[23], as well as through the temperature dependence of index of refraction of the constituent films[24], referred to as thermo-elastic and thermo-refractive noise, respectively. In stark contrast to Brownian noise, these "thermo-optic" noise sources add coherently and can in principle be eliminated through careful design of the layer structure of the mirror[25].

The most significant improvements in optical cavity performance have resulted from direct attempts to minimize the thermal noise, chiefly by choosing substrates with minimal damping, as well as by reducing the environmental temperature. Currently, the best room temperature reference cavity performance[11] is achieved by combining an ultra-low expansion (ULE) glass spacer ($\phi_{ULE} = 1.7 \times 10^{-4}$) [12] with fused silica substrates ($\phi_{silica} = 10^{-6}$-$10^{-7}$) [12,23]. Such a structure is limited primarily by the dielectric multilayer, with $\phi_{dielectric} \sim 4 \times 10^{-4}$ [12]. In cryogenic systems, for example the recently demonstrated single-crystal silicon cavity (spacer and substrates, $\phi_{Si} = 10^{-8}$ at 124 K), the coating Brownian noise becomes the sole limiting factor in the laser frequency stability[10]. In an entirely different operating regime, gravitational wave detectors are limited in their most sensitive measurement band by the coating thermal noise introduced by the lossy dielectric multilayers[22]. It is fascinating to realize that the performance of a 4-km long interferometer is limited entirely by a ~5 μm-thick surface coating.

Progress in minimizing the excessively large coating loss angle, unfortunately, has been modest. The best reported value to date for IBS-deposited dielectric multilayers is $2\times10^{-4}$ through the incorporation of a small fraction of $TiO_2$ in the high-index tantala layers[14]. Even with this improvement, the large loss angle will still remain the dominant noise source for reference cavities and gravitational wave detectors. For this reason, interferometer designs have employed workarounds that minimize the overall sensitivity to thermal noise. This can be achieved for example by lengthening the cavity[11], as fractional fluctuations in the center frequency scale inversely with the total optical path length; by increasing the size of the optical mode[26], as sampling a larger area of the mirror surface effectively averages out small-scale fluctuations; or by exploiting the coherent character of the underlying displacements and strains for potential cancellation[27]. Unfortunately, these workarounds either significantly complicate the system design or lead to excessively large cavities with increased vibration sensitivity. In a similar vein, various proposals have been put forward to significantly alter or even eliminate the coating entirely, including resonant waveguide grating reflectors[28], photonic crystal reflectors[29], ring cavities, and WGM resonators[30]. These approaches, however, have not yet demonstrated both a sufficiently high mechanical quality and sufficiently narrow cavity linewidth. Ultimately, the next generation of advanced interferometric systems will only be possible through solutions that address the mechanical properties of the coating itself.

Recent work in quantum optomechanics has provided a completely different route for a solution. This field of research exploits optomechanical interactions within optical cavities in order to control and study the quantum regime of nano- to macro-scale mechanical oscillators[16]. Similar to the requirements discussed here, reaching the quantum regime of mechanical motion necessitates the implementation of structures with both high optical and mechanical quality. In this context, monocrystalline $Al_xGa_{1-x}As$ heterostructures (AlGaAs) have been identified as a promising option for multilayer mirrors, in particular, since this materials system is capable of significantly reduced loss angles[31,32]. While AlGaAs-based epitaxial distributed Bragg reflectors

(DBRs) have been applied for the fabrication of vertical-cavity surface-emitting lasers (VCSELs) since the late 1980s, the mechanical damping had not been studied in detail until recently. Ring-down measurements on free-standing mechanical resonators microfabricated directly from epitaxial AlGaAs multilayers have yielded exceptional quality factors, $Q$, up to 40,000 at room temperature[32]. For an isolated resonance, $Q$ can be converted to loss angle via $\phi_{AlGaAs} = Q^{-1}$ ~$2.5\times10^{-5}$. In comparison, measured $Q$ values for free-standing $SiO_2/Ta_2O_5$ fall in the range of a few thousand[33], yielding values consistent with the coating loss angles observed in studies of such mirrors for optical reference cavities and gravitational wave detectors[19-21]. Taking into account the competitive optical performance of AlGaAs-based DBRs, with reflectivity values routinely surpassing 0.9998 [31,32], these epitaxial multilayers represent a promising alternative for the development of ultra-low thermal noise mirror structures.

While excellent optomechanical properties have been demonstrated in suspended micrometer-scale resonators, it is not immediately clear as to how these epitaxial multilayers can be deployed as high-performance macroscopic mirrors. In particular, lattice matching constraints preclude the deposition of monocrystalline films onto amorphous substrates due to the lack of a crystalline template for seeded growth. An additional difficulty arises as stable optical cavities require curved mirrors, the realization of which is incompatible with the current capabilities of high-quality epitaxial film growth. Such limitations are not found with sputtered dielectric mirrors, which can be deposited on essentially any relevant, and even structured, substrate. Nonetheless, by exploiting advanced semiconductor microfabrication processes, we have realized a successful implementation of this low-loss materials system in a standard optical reference cavity configuration. To generate the crystalline coatings, we employ a direct-bonding process that builds upon foundational work such as semiconductor "wafer fusion"[34], which has enabled the production of various engineered substrates (e.g. bonded silicon-on-insulator), as well as early demonstrations of epitaxial layer transfer[35] and more recent stamp-mediated methods[17] aimed at the development of novel micro- and optoelectronic systems. At its most basic level, these processes are analogous to optical contacting, a widely used bonding approach that is routinely employed in the construction of optical subassemblies.

Our fabrication process begins with the growth of a high quality epitaxial Bragg mirror on a GaAs substrate (Fig. 1). The multilayer used in this experiment (Fig. 1a) is a standard quarter-wave design, with alternating high and low index layers generated by modulating the Al content of the constituent films. Employing molecular beam epitaxy (MBE), we deposit high-quality epilayers (Fig. 1b) exhibiting a low-transmission stopband roughly centered on 1064 nm (Fig. 1c). As described in the Methods Summary, the substrate-transfer process entails removing the epitaxial layers from the native GaAs growth wafer, followed by direct bonding to the final host substrate (Fig. 2a). Using this technique, we realize high-quality compound-semiconductor-based multilayers transferred to planar and curved super-polished fused silica substrates. A photograph of a completed mirror assembly with a 1-m radius of curvature (ROC) is presented in Figure 2b. Two of these mirrors, employing crystalline coatings as the reflective elements, are optically contacted to a rigid 35-mm long zerodur spacer in order to construct a Fabry-Pérot reference cavity (Fig. 2c.). Measurement of the optical field ringdown of this resonator yields a cavity finesse of $1.5(1) \times 10^5$ (Fig. 2d). This result is in excellent agreement with the theoretical estimate of $1.53 \times 10^5$ based on independent measurements of the transmission (4 ppm), scatter (4 ppm), and absorption (12.5 ppm) loss and is, to our knowledge, the highest finesse value ever reported for a cavity employing dual single-crystal semiconductor mirrors. The exemplary optical properties observed here are attributed to the smooth interfaces and low background doping possible with an optimized MBE process. Measurement of the mode structure of the cavity reveals relatively strong birefringence, with two distinct polarization eigenmodes separated by 4.0(4) MHz, much larger than the cavity linewidth of 29 kHz. This value roughly matches a calculated maximum birefringence of 5.3 MHz (see Supplementary Information), with the most likely origin being intrinsic strain in the epitaxial films due to differential thermal expansion of the high and low index layers.

With the optical properties verified, the ultimate test is to probe the mechanical damping of our substrate-transferred crystalline coatings. It is a reasonable concern that the high mechanical *Q* values measured for suspended structures might be degraded by excess loss at the bonded interface. Thus, we perform direct measurement of the thermal noise through the construction of a cavity-stabilized laser at 1064 nm. The setup consists of a solid-state Nd:YAG laser precisely frequency stabilized to our crystalline-coating cavity. The short cavity length (35 mm) is chosen to make Brownian noise a dominant contribution to the overall cavity instability. To minimize noise from mechanical vibrations, the cavity is held vertically at its midplane by a ring resting on three Teflon rods[7,36] in a temperature controlled vacuum chamber (Fig. 2 and 3).

Evaluation of the cavity noise properties is realized by comparing the cavity-stabilized laser with a state-of-the-art narrow-linewidth 698 nm Sr lattice clock laser[11], which exhibits a record stability of $1\times10^{-16}$ from 1 to 1000 s (Fig. 3). Given the disparate operating wavelengths of the sources, we employ an Yb fiber comb to transfer the frequency stability between the reference at 698 nm and the system under test at 1064 nm [37]. The Yb comb output, centered at 1050 nm with a 50 nm wavelength span, is amplified and spectrally-broadened by a highly nonlinear fiber, resulting in a nearly flat and octave-spanning supercontinuum from approximately 700 nm to 1400 nm. With the Yb comb phase-locked to the clock laser, the coherence of the optical reference is distributed through the entire comb spectrum. Finally, a heterodyne beat between an individual comb line and the stabilized laser at 1064 nm, $f_{b1064}$, is detected and analysed, yielding the noise performance of the crystalline-coating cavity.

Figure 4a shows the fractional frequency stability $\sigma_y$ (Allan deviation) of $f_{b1064}$ measured with a frequency counter. After accounting for linear drifts in the center frequency of the cavity-stabilized laser, the FWHM linewidth of $f_{b1064}$ is found to be 0.7 Hz (Fig. 4a inset). With the second-order drift removed, the typical Allan deviation is $\sigma_y = 1.05(1)\times10^{-15}$ at 1 s. In Figure 4b (gray curve) we present the measured NPSD of the crystalline-coating-cavity stabilized laser.

The detailed thermal noise calculation for this cavity is provided in the Methods Summary and Supplementary Information. The Allan deviation and NPSD measurements are fully consistent with each other and with the predicted thermal noise floor. The calculated NPSD in the range of 1 Hz to 100 Hz is shown as the violet line in Figure 4b, which includes the Brownian noise contributions from the spacer, substrate and coating (together shown as the dotted line in Fig. 4b), as well as the substrate thermo-elastic noise and the coating thermo-optic noise. Each of these contributions is detailed in Table A of the Methods Summary.

Using the known damping values for the cavity materials, we extract a loss angle of $0(4) \times 10^{-5}$ for the substrate-transferred crystalline coatings, in general agreement with the measured room temperature loss angle of $2.5 \times 10^{-5}$ for low-frequency AlGaAs-based micromechanical resonators[32,a]. These results confirm that the high $Q$ value of the crystalline multilayer is maintained after the bonding process, enabling the implementation of such high-performance coatings in next generation ultra-low thermal noise optical interferometers. To summarize and stress the key result of the present work, Figure 4c highlights the thermal-noise contribution of the crystalline coating itself, comprising the Brownian and thermo-optic noise of the AlGaAs multilayer. Compared with the noise of a typical amorphous $SiO_2/Ta_2O_5$ coating, we observe at least a factor 10 reduction in the coating noise at 1 Hz. Due to the additional contribution of the AlGaAs thermo-optic noise, the crystalline coating noise is nearly equal to that of $SiO_2/Ta_2O_5$ at 100 Hz. However, the excess thermo-optic noise found with our current multilayer design can be reduced by two orders of magnitude through the addition of a half-wavelength cap (see Figure S3, Supplementary Information). In that case, the crystalline coating will be limited solely by the coating Brownian noise.

---

[a] An identical calculation assuming a fused silica $Q$ of $10^7$, consistent with the best reported measurement for this material, yields $\phi_{AlGaAs} = 4(4) \times 10^{-5}$, still in agreement with the free-standing resonator loss angle.

With the introduction of high-reflectivity end mirrors based on bonded epitaxial AlGaAs DBRs, we have demonstrated unprecedentedly low Brownian noise in an optical interferometer. The observed tenfold reduction in coating loss angle compared with state-of-the-art IBS-deposited multilayers represents a long-awaited breakthrough for the precision measurement community. Combining our crystalline coatings with optimized cavity designs will result in an immediate enhancement of the achievable frequency stability of ultra-narrow-linewidth lasers, opening up the $10^{-17}$ stability regime at room temperature. Furthermore, the reduced thermal noise in optical coatings provides a path towards ultrastable, compact, and portable laser systems and optical atomic clocks.

We anticipate immediate improvements in the optomechanical performance of these structures. As opposed to dielectric multilayers where the $Q$ saturates or even decreases at low temperatures[38], measurements of free-standing epitaxial multilayers in a cryogenic environment reveal loss angles nearly another order of magnitude lower, down to $4.5 \times 10^{-6}$ ($Q$ of $2.2 \times 10^5$) at 10 K [32], promising substantial improvements in the thermal noise performance of next-generation cryogenic cavities. Additionally, with our microfabrication-based substrate transfer process, we foresee no fundamental barriers in realizing mirror sizes relevant for interferometric gravitational wave detectors. Finally, via bandgap engineering or simple thickness variation, as discussed in the Supplementary Information, high reflectivity AlGaAs DBRs can cover the wavelength range from 650 nm, where band-edge absorption becomes problematic, to ~3 μm, with the long-wavelength limit being free-carrier absorption.

**Figures and Captions:**

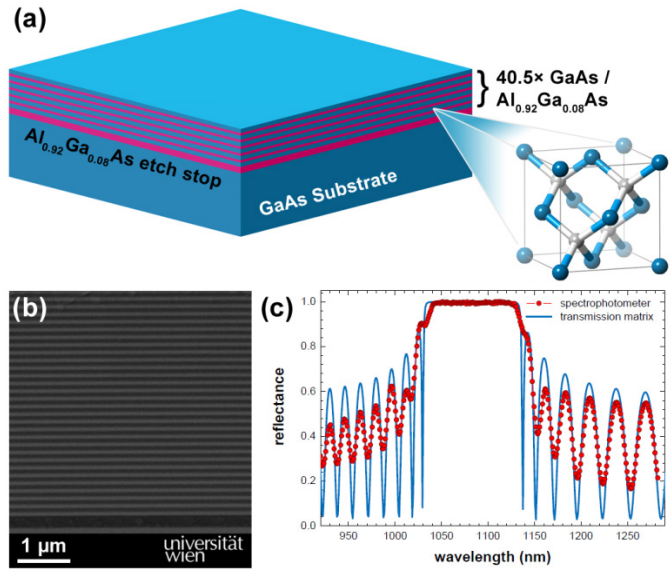

Figure 1: Material structure and reflectivity of the AlGaAs distributed Bragg reflector (DBR). a) Cross-sectional schematic of the crystalline multilayer design with the inset showing the zincblende unit cell (space group $T_d^2$-$F\bar{4}3m$). The mirror consists of 40.5 periods (81 total layers, 6.83 µm in thickness) of alternating quarter-wave GaAs (high index, 3.480 at 1064 nm and 300 K) and $Al_{0.92}Ga_{0.08}As$ (low index, 2.977 at 1064 nm and 300 K) grown on a (100)-oriented GaAs substrate via MBE. The base of the mirror incorporates a thick (270 nm) $Al_{0.92}Ga_{0.08}As$ etch stop layer for backside protection during the substrate removal process. b) Scanning electron micrograph of a cleaved facet of the epitaxial structure, demonstrating smooth and abrupt interfaces. The etch stop can be seen as the thick dark band near the bottom of the image. c) Fitted reflectance spectrum (measurement: red points, transmission matrix theory: blue line) of the AlGaAs multilayer after transfer to a glass substrate.

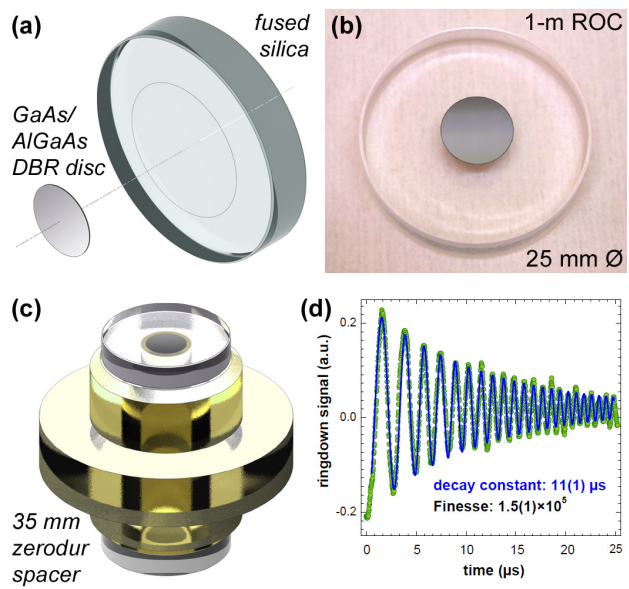

Figure 2: Construction of an optical reference cavity using substrate-transferred crystalline coatings. a) Exploded view of a bonded mirror assembly, showing the released AlGaAs mirror disc and fused silica substrate with a polish-imparted 1-m radius of curvature (ROC). b) Photograph of the front face of a completed curved mirror assembly incorporating a directly-bonded 8-mm diameter epitaxial DBR on a 25 mm diameter fused silica substrate with a 1-m ROC. This sample shows a clean and defect-free bond interface with a lack of Newton's rings between the AlGaAs and fused silica. c) Solid model of a cavity constructed from two bonded mirror assemblies (one planar and one with a 1-m ROC) optically contacted to a 35-mm long zerodur spacer. The substrate backside anti-reflection coating has been omitted for clarity. d) Swept laser frequency ringdown of the optical cavity field in reflection, yielding a finesse of $1.5(1) \times 10^5$ at 1064 nm (measurement: green points, theoretical fit: blue line) and demonstrating the excellent optical properties of the bonded crystalline coating.

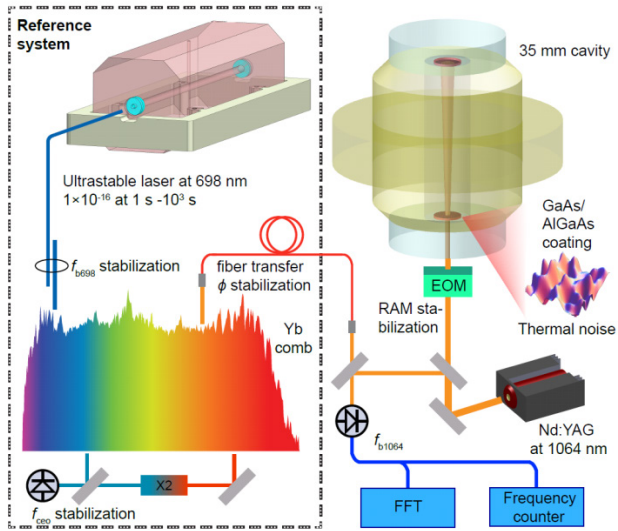

Figure 3: Thermal noise measurement system for the characterization of the crystalline-coating cavity. A narrow linewidth 698 nm laser, stabilized to a separate state-of-the-art optical cavity with a frequency stability of $1\times10^{-16}$ from 1 to 1000 s [11] is used as an optical reference and compared with a Nd:YAG laser at 1064 nm stabilized to the crystalline-coating cavity. Given the large frequency difference between the lasers, a self-referenced Yb-doped fiber femtosecond frequency comb is used to transfer the stability of the 698 nm reference to 1064 nm in order to facilitate a direct optical heterodyne beat comparison between the two laser/cavity systems. All fiber links employed in the setup are phase stabilized. The residual amplitude modulation of the electro-optic modulator (EOM) used for the generation of cavity servo error is actively stabilized to $3.5\times10^{-16}$ at 1 s. The beat signal ($f_{b1064}$) is analyzed with a fast-Fourier-transform (FFT) spectrum analyzer and a frequency counter, yielding the noise properties of the cavity and crystalline coatings.

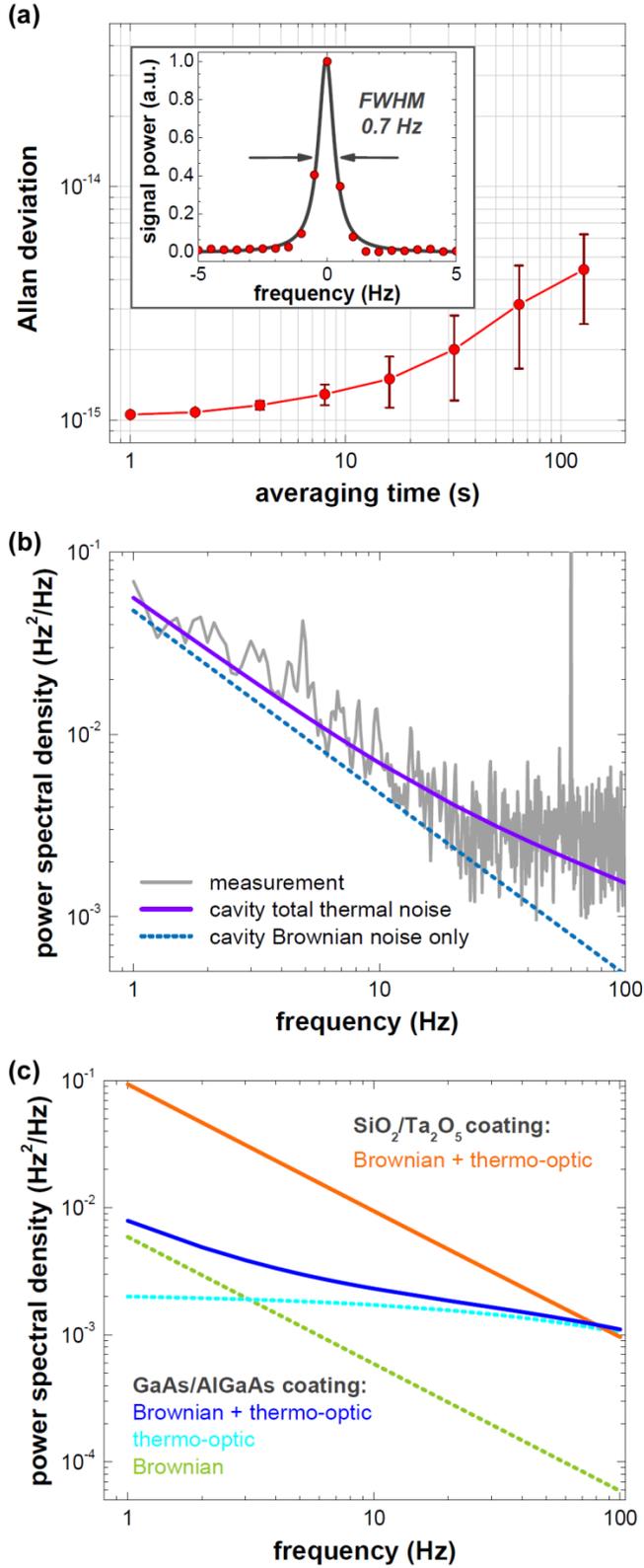

Figure 4: Characterization of the crystalline-coating-stabilized 1064 nm laser noise performance. a) Allan deviation between 1 and ~100 s. The measured $f_{b1064}$ (red data points, main plot) yields $\sigma_y$ of $1.05(1)\times10^{-15}$ at 1 s and is limited by higher-order laser frequency drift beyond 10 s. Inset, measured laser linewidth near 1064 nm. The Lorentzian fit (dark gray line) yields a linewidth of 0.7 Hz (for a resolution bandwidth of 0.5 Hz). b) The stabilized laser frequency noise power spectral density, NPSD (1 to 100 Hz), with thermal-noise-limited performance (gray curve). Violet line: total thermal noise of the reference cavity based on a summation of the Brownian noise (coating, substrate, and spacer), thermo-optic noise (coating), and thermo-elastic noise (substrate) contributions, assuming $\phi_{spacer}$ is $3\times10^{-4}$, $\phi_{sub}$ is $1\times10^{-6}$, $\phi_{AlGaAs}$ is $2.5\times10^{-5}$. Dashed-turquoise: cavity Brownian noise with contributions from the coating, substrate, and spacer. c) Comparison of the coating thermal noise for a conventional dielectric mirror design (20 periods of alternating $SiO_2/Ta_2O_5$ with a half-wavelength $SiO_2$ cap) and our AlGaAs-based crystalline coatings. The orange line is the thermal noise of the dielectric multilayer including both Brownian and thermo-optic contributions, while the thick blue line indicates the thermal noise of the crystalline coating. This curve incorporates both the Brownian noise (green) and the thermo-optic noise (cyan).

## Methods Summary:

*Mirror fabrication*: Our novel substrate-transfer coating procedure entails separating the epitaxial multilayer from the original growth wafer and directly bonding it—using no adhesives or intermediate films—to the desired mirror blank. Thus, the bonded mirror assembly initially begins as two separate components, a GaAs wafer capped with an epitaxial DBR and a super polished fused-silica substrate with a standard backside anti-reflection coating. The DBR is grown via MBE on a 150-mm diameter semi-insulating GaAs wafer and comprises alternating quarter-wave GaAs for the high index layers and $Al_{0.92}Ga_{0.08}As$ for the low index films. Generation of the AlGaAs mirror disc relies on standard microfabrication steps including optical lithography, in order to define the lateral geometry of the disc, followed by chemical etching to extrude the disc shape through the epitaxial multilayer. Chemo-mechanical substrate removal employing lapping, followed by selective wet chemical etching, is used to strip the GaAs growth template. Next, the thick AlGaAs etch stop layer (Figs. 1a and 1b) is removed and the mirror surface is cleaned of any potential debris. Finally, the crystalline mirror disc and the silica substrate are pressed into contact, resulting in a spontaneous van der Waals bond. To strengthen the interface and minimize potential frictional losses, a post-bond anneal completes the fabrication procedure.

*Thermal noise theory*: The power spectral density of the thermal noise displacement ($G_{total}$), consisting of Brownian motion of the spacer ($G_{spacer}$), substrate ($G_{sub}$), and coating ($G_c$), as well as substrate thermo-elastic noise ($G_{TE}$) and coating thermo-optic noise ($G_{TO}$), is given by:

$$G_{total}(f) = G_{spacer}(f) + G_{sub}(f) + G_c(f) + G_{TE}(f) + G_{TO}(f)$$

$$= \frac{2k_BT}{\pi^2 f} \frac{L}{(R^2-r^2)Y_{spacer}} \phi_{spacer} + \frac{4k_BT}{\pi^{3/2} f} \frac{1-\sigma_{sub}^2}{w_m Y_{sub}} \phi_{sub}$$

$$+ \frac{4k_BT}{\pi^2 f} \frac{1-\sigma_{sub}^2}{w_m Y_{sub}} \frac{D}{w_m} \frac{\phi_c}{Y_{sub}Y_c(1-\sigma_c^2)(1-\sigma_{sub}^2)} \left[Y_c^2(1+\sigma_{sub})^2(1-2\sigma_{sub})^2 + Y_{sub}^2(1+\sigma_c)^2(1-2\sigma_c)\right]$$

$$+ G_{TE}(f) + G_{TO}(f) \quad (1)$$

Here,

$$G_{TE}(f) = \frac{8}{\sqrt{\pi}}(1+\sigma_{sub})^2 \alpha_{sub}^2 \frac{k_BT^2 w_m}{\kappa_{sub}} \times \int_0^{+\infty} du \int_{-\infty}^{+\infty} dv \sqrt{2/\pi^3} u^3 e^{-u^2/2} / \{(u^2+v^2)[(u^2+v^2)^2 + \Omega_{TE}^2(f)]\}$$

$$\Omega_{TE}(f) = w_m^2 C_{sub} \pi f / \kappa_{sub} ,$$

and

$$G_{TO}(f) = \frac{4k_B T^2}{\pi^{3/2} w_m^2 \sqrt{f \kappa_c C_c}} \Gamma_{TO}(f)(\bar{\alpha}_c D - \bar{\beta}_c \lambda - \bar{\alpha}_{sub} D C_c / C_{sub})^2 \qquad (w_m^2 C_c \pi f / \kappa_c \gg 1)$$

$$G_{TO}(f) = \frac{4k_B T^2}{\sqrt{\pi} w_m \kappa_c} \Gamma_{TO}(f)(\bar{\alpha}_c D - \bar{\beta}_c \lambda - \bar{\alpha}_{sub} D C_c / C_{sub})^2 \qquad (w_m^2 C_c \pi f / \kappa_c \ll 1).$$

All symbols and their relevant values have been defined in the Supplementary Information.

The corresponding frequency noise power spectra density $G_\nu$ is

$$G_\nu(f) = G_x(f) \times \frac{\nu^2}{L^2}. \qquad (2)$$

Here $\nu$ is optical frequency and $L$ the cavity length. Eqs. 1 and 2 are used to calculate the thermal-noise-limited NPSD, shown as the solid violet line in Figure 4b.

The corresponding thermal noise-limited fractional frequency stability, $\sigma_y$, can be determined from

$$\sigma_y^2(\tau) = \int_0^\infty \frac{G_\nu(f)}{\nu^2} \times 32 \frac{(\sin(\pi f \tau/2))^4 \times |\sin(\pi f \tau)|^2}{(\pi f \tau)^4} df. \qquad (3)$$

The values of $G_x$ at 1 Hz and $\sigma_y$ at 1 s are summarized in Table A. For our current cavity configuration, Brownian noise represents the dominant contribution to the predicted total thermal noise, which gives rise to a $\sigma_y$ of ~1.08×10$^{-15}$ at 1 s. The contribution of the coating Browning noise is large enough (11% as shown in Table A below) to be isolated from competing noise sources, allowing us to determine the coating loss angle.

**Table A**. Theoretical calculation of thermal noise. Displacement values $G_x$ are shown at 1 Hz and the Allan deviation $\sigma_y$ at 1 s. The assumed loss angles of zerodur, fused silica, and coating are 3×10$^{-4}$, 1×10$^{-6}$, and 2.5×10$^{-5}$, respectively.

| Noise type | | $G_x$ (×10$^{-34}$) m$^2$/Hz | $\sigma_y$ (×10$^{-16}$) | Contribution to total ($\sigma_y$)$^2$ | | Total $\sigma_y$ (×10$^{-15}$) |
|---|---|---|---|---|---|---|
| Brownian motion | Spacer | 4.83 | 8.4 | 60% | 91% | 1.08 |
| | Substrate | 1.60 | 4.8 | 20% | | |
| | Coating | 0.91 | 3.6 | 11% | | |
| Substrate thermoelastic | | 0.98 | 3.2 | 8.8% | | |
| Coating thermo-optic | | 0.045 | 0.5 | 0.2% | | |

**Acknowledgments** We wish to thank R. X. Adhikari, A. Alexandrovski, C. Benko, G. M. Harry, R. Lalezari, L.-S. Ma, E. Murphy, M. Notcutt, S. D. Penn, A. Peters, P. Ullmann, and R. Yanka for stimulating discussions and technical assistance. Work at the University of Vienna is supported by the Austrian Science Fund (FWF), the European Commission, and the European Research Council (ERC) Starting Grant Program. The work at CMS is supported by the Austria Wirtschaftsservice GmbH (AWS) and the ERC Proof of Concept Initiative. Work at JILA is supported by the US National Institute of Standards and Technology (NIST), the DARPA QuASAR Program, and the US National Science Foundation (NSF) Physics Frontier Center at JILA. Microfabrication was carried out at the Center for Micro- and Nanostructures (ZMNS) of the Vienna University of Technology.

# Supplementary Information

# Tenfold reduction of Brownian noise in optical interferometry

Garrett D. Cole[1,2,*], Wei Zhang[3,*], Michael J. Martin[3], Jun Ye[3], and Markus Aspelmeyer[1]

[1] *Vienna Center for Quantum Science and Technology (VCQ), Faculty of Physics, University of Vienna, A-1090 Vienna, Austria*

[2] *Crystalline Mirror Solutions GmbH, A-1090 Vienna, Austria*

[3] *JILA, National Institute of Standards and Technology and University of Colorado, Boulder, Colorado 80309-0440, USA*

## *Finesse Estimation*

Calculation of the cavity finesse requires quantification of the various loss mechanisms of the end mirrors. As our crystalline coatings are constructed from identical multilayers (sourced from neighbouring locations on the same epitaxial wafer) we assume that the individual coatings have a matched reflectance. Below we provide a description of the steps taken in order to characterize the transmission, scatter, and absorption loss of the mirrors. In each case measurements are performed both on the original growth wafer and after bonding in order to investigate any potential changes in the optical properties following the substrate-transfer process.

---

[*] These authors contributed equally to this work.

In order to determine the mirror transmission, the reflectance spectrum of the DBR is recorded via spectrophotometry, as shown in Fig. 1c. Fitting of the mirror stopband with a transmission matrix model, using x-ray diffraction derived composition and thickness values as a baseline, it is possible to extract the individual layer thicknesses for the high and low index components of the multilayer. Fortunately, given the long history of application in optoelectronics (e.g. Ref. 1), the refractive index and dispersion characteristics are very well known for the AlGaAs materials system. This fitting technique is exquisitely sensitive to variations in thickness, with changes on the order of 1 nm for individual layer thicknesses resulting in shifts of the center wavelength of the high-reflectivity stopband by nearly 10 nm. The fitted transmission matrix model is then used for an accurate estimate of the transmission loss of the mirror. For the 40.5 period DBR employed here, the extracted transmission is 4 ppm. Independent measurement of the transmission of a post-bonded crystalline coating (on a fused silica substrate) yields 3.8 ppm, verifying the accuracy of this approach. One current limitation in our process is the lack of an *in situ* optical monitor in the MBE system. Thus, it is difficult to reliably realize a desired transmission value. With this in mind, early mirrors were designed with minimal transmittance in order to avoid the possibility of insufficient finesse.

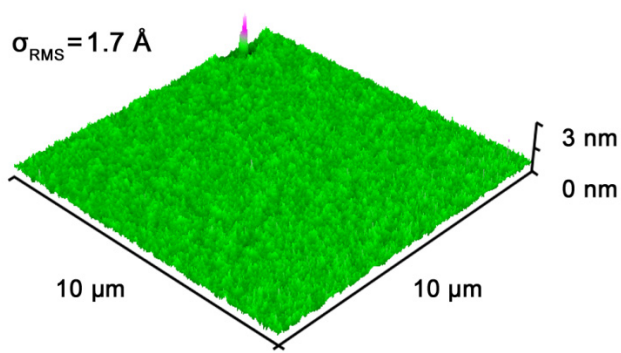

Figure S1: Post-bond AFM characterization of the AlGaAs DBR surface quality. The measured RMS roughness of 1.7 Å (determined by averaging line scans over a 10 μm × 10 μm area) leads to a calculated scatter loss of 4 ppm for the crystalline coating.

Atomic force microscopy (AFM) is used to characterize the surface roughness, and ultimately the scatter loss, of both the pre- and post-bonded mirror structure. For our MBE-grown AlGaAs DBR, we record a root mean square (RMS) roughness value of 1.7 Å over a 10 µm × 10 µm scan area, as shown in Figure S1. Note that this measurement was made following the transfer process and remains unchanged when compared with the on-wafer surface quality. Assuming normal-incidence illumination of a surface characterized by an RMS roughness $\sigma$, the scatter-limited reflectance is determined from the expression derived by Davies[2] $R_s = e^{\frac{-(4\pi\sigma)^2}{\lambda^2}}$. In the long wavelength (for surface features much smaller than the wavelength $\lambda$) and high reflectivity limit, this expression yields a fairly accurate estimate of the reflectance and thus the corresponding scatter loss value (1-$R_s$) [3]. Using our measured AFM data and a wavelength of 1064 nm, we calculate a scatter loss of 4 ppm for the AlGaAs DBR. The validity of the further simplification of taking into account only the *surface* roughness of the multilayer is ultimately verified by the excellent agreement between our theoretical and measured finesse values (1.53×10$^5$ versus 1.5(1)×10$^5$ respectively).

The final component necessary to estimate the cavity finesse is the optical absorption. This parameter is probed directly via photothermal common path interferometry (PCI) [4]. Absorption data extracted by PCI reveal typical loss levels of ~10 ppm (unchanged for both pre- and post-bond conditions) for our MBE-grown AlGaAs mirrors at 1064 nm [5], with the current best measurement below 5 ppm. For the sample used in the study, the average absorption value was slightly higher at 12.5±1.19 ppm (with a somewhat large sample-to-sample variation). A series of three linear scans (11.1±0.08 ppm, 12.9 ±0.12 ppm, and 13.5±1.18 ppm) were averaged for this measurement; a typical example is presented in Figure S2.

We are currently operating under the assumption that this property is dominated by free-carrier absorption (FCA). In contrast, calculations of the two-photon absorption at 1064 nm yields single digit or sub-ppm absorption values for realistic intra-cavity powers. Unfortunately, with these thin films and at these low carrier concentrations, direct characterization of the carrier type and concentration via Hall measurement is extremely difficult. Assuming that the corresponding absorption coefficient of 0.38 cm$^{-1}$ (calculated from the 12.5 ppm absorption value over a round-trip penetration depth of 163 nm) is dominated by a *p*-type impurity background and extrapolating from the data presented in Kudman and Seidl[6], the estimated carrier concentration for this structure is $2.5\times10^{16}$ cm$^3$.

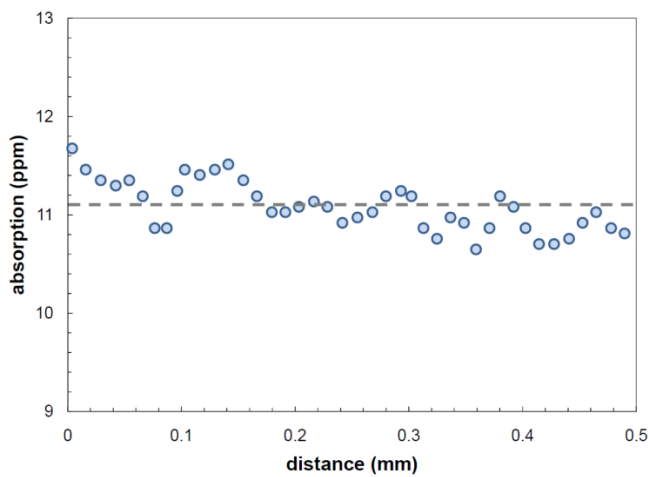

Figure S2: Typical PCI line-scan for characterizing the optical absorption of the bonded crystalline coating. The average absorption in this example is 11.1±0.08 ppm at 1064 nm. Averaging of three line scans yields a mean absorption value of 12.5 ppm for the 40.5-period AlGaAs DBR.

With FCA in the high index layers being the ultimate limit, it should be possible to generate high reflectivity AlGaAs-based mirrors out to approximately 3 µm [7], while moving to wavelengths shorter than the current implementation at 1064 nm will necessitate a significant increase in the average Al content in order to stave off interband absorption. Previous work with visible VCSELs has demonstrated high quality AlGaAs DBRs with center wavelengths as short as 650 nm [8], though further research will be necessary to quantify the minimum achievable absorption levels in this wavelength window. Note that operation at 698 nm will be directly relevant for Sr optical atomic clock applications.

*Birefringence Calculations*

As discussed in the manuscript, the mode structure of the fixed spacer cavity constructed with the AlGaAs-based crystalline coatings reveals two orthogonal polarization eigenmodes with a splitting of 4(0.4) MHz. Interestingly, the refractive index of unstrained (100)-oriented ternary $Al_xGa_{1-x}As$ alloys is in-plane isotropic, meaning that, from the point of view of a reflecting surface-normal beam, there is no radial variation in index. Thus, for an ideal cavity constructed from perfect crystalline coatings, there will be no birefringence. However, anisotropy can easily be introduced through the application of strain in the system, leading to a corresponding directional dependence on index via the elasto-optic effect. For the bonded mirrors there are two possible paths for strain being imparted on the mirror: 1) bending-induced strain generated when fabricating the curved mirror and 2) internal strains resulting from the epitaxial growth process.

Path 1 requires a finite element model in order to calculate the imparted strain on the deformed mirror. In this case the mirror disc (8 mm in diameter by 6.83 µm in thickness) is forced into a prescribed displacement described by $x^2 + y^2 + (z - 0.999992)^2 = 1$ m (with *x* and *y* lying within the plane of the originally flat mirror and *z* being the vertical direction normal to the DBR surface), forcing the mirror to conform to a sphere of 1-m radius with a maximum deflection of 8 µm at the mirror center. Incorporating the average density and elastic constants of the single-crystal multilayer ($\rho$=4551 kg/m$^3$; $c_{11}$=119.94, $c_{12}$=55.38, and $c_{44}$=59.15 GPa), the mean in-plane stress, $\frac{1}{2}(\sigma_{xx} + \sigma_{yy})$, is found to be approximately 500 kPa (compressive) on the front surface of the mirror and is essentially uniform across the mirror face given the constant radius of curvature of the deformed multilayer. Note that this is an extremely small stress value compared with the hundreds of MPa or even GPa levels of internal stress found in typical ion beam sputtered films. There are two take-home messages from this exercise: i) for a bonded mirror with a 1-m ROC, we are imparting very little strain in the multilayer, so there is no fear of damaging the structure, and ii) given the miniscule stress (and corresponding strain value), this is likely not the cause of the birefringence.

For a quick estimate, path 2 (growth-induced strain) can be calculated using simple analytical expressions. As we employ a well-known epitaxial materials system, it is quite easy to extract the internal strain arising from slight differences in the inter-atomic spacing between the constituent components of the multilayer, or in other words, due to the differential thermal contraction of the layers upon cooling, from the growth, to room temperature, $T$. Notably, for $Al_xGa_{1-x}As$ compounds, the lattice parameter is nearly exactly matched for typical deposition temperatures[9]. At room temperature, the high index film, consisting of binary GaAs, is characterized by a lattice parameter, $a$, of 5.6533 Å, while for $Al_{0.92}Ga_{0.08}As$ this value is 5.6605 Å. This slight difference in lattice constant leads to a mismatch strain, $e_m$, of $1.27\times10^{-3}$ when cooling from ~900 K to 300 K, in general agreement with previous investigations showing CTE-driven mismatch values slightly over 1 millistrain for films of a similar average aluminum content[10-12]. Assuming an average isotropic Young's modulus, $E$, of 100 GPa for the multilayer, the resulting mismatch-driven stress, $X$, is then 127 MPa (simply determined from $\sigma = eE$). Obviously, this value is significantly larger (>250×) than the bending-induced stress described above and is thus the more likely source of the observed polarization splitting.

Realizing a first-order estimate of the index change with this internal strain is fairly straightforward. The principal axes of birefringence in zincblende crystals under a uniform applied strain is the [110] family of directions; in that case the resulting index difference[13] between the parallel and perpendicular axes is described by $\Delta n = \frac{1}{2}\sqrt{\varepsilon}^{-1}\alpha_{pe}X$, where ε is the real part of the dielectric constant at optical frequencies, $\alpha_{pe}$ is the photoelastic coefficient, and X is the internal stress in the multilayer. Plugging in the average photoelastic coefficient of $5\times10^{-11}$ Pa$^{-1}$ and high frequency dielectric constant of 12, both values from Ref. [12], along with the previously calculated stress value (127 MPa), yields an index difference of roughly $1\times10^{-3}$.

Calculation of the frequency splitting requires knowledge of the optical path length including the vacuum cavity ($L_{cav}$ = 35 mm), plus the additional field penetration depth into the mirrors[14]. For the 40.5-period AlGaAs DBR employed in this experiment, the penetration depth, $L_{pen}$, at 1064 nm is calculated to be 163 nm. The change in length from the strain-driven index perturbation, δ, is determined to be $2\,L_{pen}\,\Delta n$ = 3.3 Å. Converting this physical length to a frequency difference, Δf, results in a maximum polarization splitting of 5.3 MHz for a center wavelength of 1064 nm ($f_0$=2.82×10$^{14}$ Hz) using $\Delta f \approx f_0 \frac{2\delta}{L_{cav}}$. This value compares well with our measurement of 4(0.4) MHz for a randomly oriented pair of crystalline coatings. For future runs, the mirror discs should include features indicating the crystallographic orientation of the coatings in order to maintain the desired alignment.

## *Thermal Noise Measurement and Calculations*

The Methods Summary provides basic information as required for the calculation of the thermal noise results presented in the Letter. Below we present a detailed discussion of the thermal noise model and introduce the relevant parameters used in the present work.

**Thermal noise theory:** The power spectral density of the thermal noise displacement ($G_{total}$), consisting of Brownian motion of the spacer ($G_{spacer}$), substrate ($G_{sub}$), and coating ($G_c$), substrate thermoelastic noise ($G_{TE}$) and coating thermo-optic noise ($G_{TO}$), is described by the following[5,15-21] (symbols and corresponding values defined in Table 1):

$$G_{total}(f) = G_{spacer}(f) + G_{sub}(f) + G_c(f) + G_{TE}(f) + G_{TO}(f)$$
$$= \frac{2k_B T}{\pi^2 f} \frac{L}{(R^2 - r^2) Y_{spacer}} \phi_{spacer} + \frac{4k_B T}{\pi^{3/2} f} \frac{1-\sigma_{sub}^2}{w_m Y_{sub}} \phi_{sub}$$
$$+ \frac{4k_B T}{\pi^2 f} \frac{1-\sigma_{sub}^2}{w_m Y_{sub}} \frac{D}{w_m} \frac{\phi_c}{Y_{sub} Y_c (1-\sigma_c^2)(1-\sigma_{sub}^2)} \left[ Y_c^2 (1+\sigma_{sub})^2 (1-2\sigma_{sub})^2 + Y_{sub}^2 (1+\sigma_c)^2 (1-2\sigma_c) \right]$$
$$+ G_{TE}(f) + G_{TO}(f) \qquad (1)$$

Here,

$$G_{TE}(f) = \frac{8}{\sqrt{\pi}} (1+\sigma_{sub})^2 \alpha_{sub}^2 \frac{k_B T^2 w_m}{\kappa_{sub}} \times \int_0^{+\infty} du \int_{-\infty}^{+\infty} dv \sqrt{2/\pi^3} u^3 e^{-u^2/2} / \{(u^2+v^2)[(u^2+v^2)^2 + \Omega_{TE}^2(f)]\}$$

$$\Omega_{TE}(f) = w_m^2 C_{sub} \pi f / \kappa_{sub}$$

and

$$G_{TO}(f) = \frac{4k_B T^2}{\pi^{3/2} w_m^2 \sqrt{f \kappa_c C_c}} \Gamma_{TO}(f) (\overline{\alpha}_c D - \overline{\beta}_c \lambda - \overline{\alpha}_{sub} D C_c / C_{sub})^2 \qquad (w_m^2 C_c \pi f / \kappa_c \gg 1)$$

$$G_{TO}(f) = \frac{4k_B T^2}{\sqrt{\pi} w_m \kappa_c} \Gamma_{TO}(f) (\overline{\alpha}_c D - \overline{\beta}_c \lambda - \overline{\alpha}_{sub} D C_c / C_{sub})^2 \qquad (w_m^2 C_c \pi f / \kappa_c \ll 1),$$

where $\Gamma_{TO}(f)$ is the thickness correction factor of the coating layer[5,21]. The expression of $G_{TO}$ at high and low frequencies are based on the treatments shown in Ref. 20 and 21.

**Table 1**. Symbols and parameters used in the thermal noise calculations. Here, "*spacer*" refers to the zerodur cavity spacer, the "*sub*" index is for the fused silica substrates, and "*c*" represents the AlGaAs crystalline coating.

| Parameter | Description | Value |
|---|---|---|
| $\lambda$ | wavelength | 1.064 µm |
| $f$ | Fourier frequency | |
| $L$ | Cavity length | 35 mm |
| $R$ | Spacer radius | 15 mm |
| $r$ | Central bore radius | 5 mm |
| $Y_{spacer/sub/c}$ | Young's modulus spacer/substrate/coating | 91 / 72 / 100 GPa |
| $\phi_{spacer/sub/c}$ | Loss angle spacer/substrate/coating | $3\times 10^{-4}$ / $1\times 10^{-6}$ / $2.5\times 10^{-5}$ |
| $\sigma_{sub/c}$ | Poisson ratio substrate/coating | 0.17/0.32 |
| $w_m$ | Beam radius | 250 µm |
| $D$ | Coating thickness | 6.83 µm |
| $T$ | Temperature | 300 K |
| $k_B$ | Boltzmann's constant | $1.38\times 10^{-23}$ J/K |
| $C_{sub/c}$ | Heat capacity per volume substrate/coating | $1.71\times 10^{6}$ / $1.639\times 10^{6}$ J/ Km$^3$ |
| $\kappa_{sub/c}$ | Thermal conductivity substrate/coating | 1.38 / 62.9 W/Km |
| $\overline{\alpha}_{sub/c}$ | Effective coefficient of thermal expansion substrate/coating | $1.2\times 10^{-6}$ / $1.68\times 10^{-5}$ 1/K |
| $\overline{\beta}_c$ | Effective thermorefractive coefficient (coating) | $5\times 10^{-4}$ |

The corresponding frequency noise $G_\nu$ is

$$G_\nu(f) = G_x(f) \times \frac{\nu^2}{L^2}, \qquad (2)$$

where $\nu$ is the optical frequency. Eqs. 1 and 2 allow us to determine the theoretical thermal-noise-limited frequency noise, which is shown as the violet line in Figure 4b.

For the type of frequency counter employed in this measurement, the theoretical thermal noise-limited fractional frequency stability, $\sigma_y$, is equal to[22]

$$\sigma_y^2(\tau) = \int_0^\infty \frac{G_\nu(f)}{\nu^2} \times 32 \frac{(\sin(\pi f \tau/2))^4 \times |\sin(\pi f \tau)|^2}{(\pi f \tau)^4} df. \qquad (3)$$

Employing Eqs. 1 to 3 and the parameters of the Fabry-Pérot cavity listed in Table 1, the thermal noise limited values of $G_x$ and $\sigma_y$ (at 1 Hz and 1 s, respectively) can be calculated. These results are summarized in Table 2. To facilitate measurement of the contribution from the mirror coatings, a relatively short spacer length of 35 mm is chosen for the experiment.

**Table** 2. Theoretical calculation of thermal noise in the crystalline-coating cavity. Displacement values $G_x$ are shown at 1 Hz and the Allan deviation $\sigma_y$ at 1 s. Assumed loss angles of zerodur, fused silica, and the crystalline coating are $3\times10^{-4}$, $1\times10^{-6}$, and $2.5\times10^{-5}$, respectively.

| Noise type | | $G_x$ (×$10^{-34}$) m²/Hz | $\sigma_y$ (× $10^{-16}$) | Contribution to total $(\sigma_y)^2$ | | Total $\sigma_y$ (× $10^{-15}$) |
|---|---|---|---|---|---|---|
| Brownian motion | Spacer | 4.83 | 8.4 | 60% | 91% | 1.08 |
| | Substrate | 1.60 | 4.8 | 20% | | |
| | Coating | 0.91 | 3.6 | 11% | | |
| Substrate thermoelastic | | 0.98 | 3.2 | 8.8% | | |
| Coating thermo-optic | | 0.045 | 0.5 | 0.2% | | |

The results in Table 2 clearly demonstrate that Brownian noise makes the main contribution to the predicted total thermal noise. The fractional frequency stability as predicted by the total theoretical thermal noise is $1.08\times10^{-15}$ at 1 s, where the contribution from the crystalline coating is approximately 11% when assuming a loss angle of $2.5\times10^{-5}$ as measured in free-standing resonators. This can be compared with conventional dielectric end mirrors where 70% of the loss is attributed to the amorphous $SiO_2/Ta_2O_5$ layers. Though admittedly small in terms of the overall fractional contribution, the coating Brownian noise contribution is large enough to be isolated from other noise sources, allowing us to extract the coating loss angle by measuring the stability of the 1064 nm cavity-stabilized laser. The crystalline coating loss angle is evaluated to be $0(4)\times10^{-5}$, corresponding to the measured $\sigma_y$ of $f_{b1064}$ at 1 s averaging time, using a loss angle of $3\times10^{-4}$ for the zerodur spacer and $1\times10^{-6}$ for the fused silica substrate[16]. The error in the AlGaAs loss angle is based on a 10% error in the measured $\sigma_y$ as well as a 20% uncertainty in the loss angles of the cavity components, including the zerodur spacer and fused silica substrates[*].

**Details on the Yb frequency comb:** The carrier-envelop-offset, $f_{ceo}$, is generated by a *f-2f* self-referencing interferometer and stabilized on a RF reference via feedback to an AOM, acting as the fast actuator, and the temperature of a fiber Bragg grating as the slow actuator[23]. The bandwidth of the $f_{ceo}$ servo loop is approximately 160 kHz. To phase lock the Yb comb to the 698 nm clock laser, the optical heterodyne beat, $f_{b698}$, between a comb tooth and the clock laser is stabilized using the intracavity EOM as a fast actuator and a cavity end-mirror mounted on a piezoelectric transducer (PZT), as well as a PZT-based fiber stretcher, as slow actuators. The bandwidth of the $f_{b698}$ servo loop is about 150 kHz.

**Stabilization of the 1064 nm laser:** The cavity-stabilized laser at 1064 nm is a continuous wave all solid-state, diode-pumped non-planar ring oscillator (NPRO). Since the free-running laser linewidth is only a few kHz—much narrower than the Fabry-Pérot cavity linewidth—an additional pre-stabilization stage can be avoided.

The reference cavity and supporting structure are installed in a vacuum chamber with a residual pressure of $10^{-6}$ torr provided by an ion pump (2 liter/s pumping speed). The temperature of the vacuum chamber is 300 K with variations smaller than 1 mK over a 24 hour period. The vacuum chamber and optical components for the RF sideband lock are located on a passive vibration isolation platform (Minus K, 0.5-1 Hz resonance frequency). The laser output is divided into two components: the majority of the optical power is sent to the Yb comb system for intercomparison, while the remaining 10% of the output power is fiber-coupled to the platform for RF sideband locking to the reference cavity. A fiber-based EOM provides phase modulation for locking, with its residual amplitude modulation (RAM) actively stabilized to well below the predicted thermal noise floor[24-26]. The single frequency laser is stabilized to the Fabry-Pérot cavity via feedback to the laser cavity PZT with a servo bandwidth of approximately 40 kHz.

---

[*] If we perform an identical calculation with a fused silica loss angle of $10^{-7}$, consistent with the lowest reported measurements for this material, we obtain a loss angle of $4(4)\times10^{-5}$ for the substrate-transferred AlGaAs multilayer.

## *Reduction of excess thermo-optic with an optimized AlGaAs coating*

As discussed in the Letter, our AlGaAs multilayer displays significant thermo-optic noise at frequencies exceeding 10 Hz. This excess noise arises from the relatively large values of the effective thermorefractive and thermal expansion coefficients for this design and materials system, as can be seen in Table 1. However, it is important to note that this limitation should be fairly straightforward to overcome. According to the thermo-optic noise expression given in Eq. 1 (from Refs. [5] and [21]), the addition of a half-wavelength cap to our current multilayer design can reduce this noise contribution by 2 orders of magnitude, as shown in Figure S3. Consequently, employing such an optimized design, the coating Brownian noise is then predicted to dominate the total coating thermal noise. Compared with the theoretical thermal noise of a traditional $SiO_2/Ta_2O_5$ multilayer as shown in Figure 4c, the corresponding improvement of the AlGaAs multilayer with the additional cap layer will be at least a factor of 10 over the entire range from 1 Hz to 100 Hz. Future implementations of AlGaAs-based crystalline coatings will thus incorporate this additional material in order to significantly reduce the thermo-optic noise contribution.

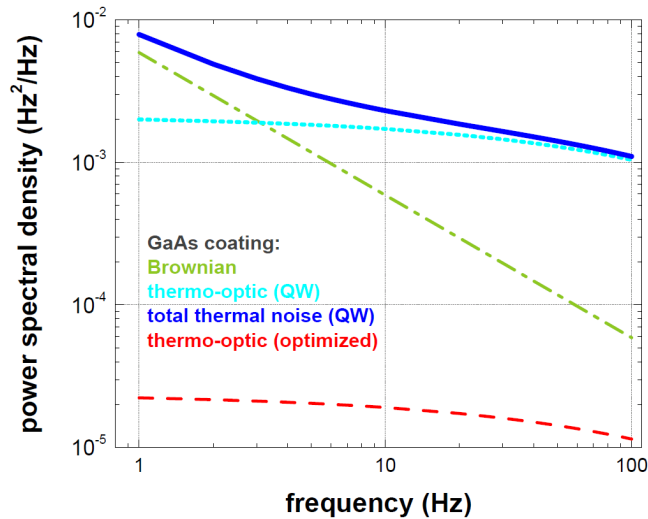

Figure S3: Comparison of the thermal noise performance of substrate-transferred AlGaAs coatings with and without a half-wavelength cap. The green dot-dashed line is the AlGaAs Brownian noise, the cyan dotted line is the thermo-optic noise for an AlGaAs multilayer using a standard quarter-wave (QW) design, while the blue solid line is the total thermal noise for such a structure. When a half wavelength cap is added to the QW AlGaAs multilayer (optimized design), the thermo-optic noise is greatly reduced, as indicated by the red dashed line. In this case the total thermal noise for the optimized structure would be limited simply by the Brownian noise curve.